\documentclass[dvips,twocolumn]{revtex4}
\usepackage{amsmath,amssymb,amsthm}
\usepackage[pdftex]{graphicx}
\usepackage{bm}
\usepackage{mathrsfs}

\begin{document}
\title{Quantumness of Gaussian Discord: Experimental Evidence and Role of System-Environment Correlations}
\author{Vanessa Chille$^{1}$,  Niall Quinn$^{2}$,   Christian Peuntinger$^{1}$, Callum Croal$^{2}$, Ladislav Mi\v{s}ta, Jr.$^{3}$, Christoph Marquardt$^{1}$, Gerd Leuchs$^{1}$, Natalia Korolkova$^{2}$}
\affiliation{$^1$Max Planck Institute for the Science of Light, G\"unther-Scharowsky-Str. 1/Bldg. 24, Erlangen, Germany\\
Institute of Optics, Information and Photonics, University of Erlangen-N\"urnberg, Staudtstr.
7/B2, Erlangen, Germany\\
$^2$School of Physics and Astronomy, University of St. Andrews, North Haugh, St. Andrews KY16 9SS, UK\\
$^3$Department of Optics, Palack\' y University, 17. listopadu 12,  771~46 Olomouc, Czech Republic
}

\begin{abstract}
We provide experimental evidence of quantum features in bi-partite states classified as entirely classical according to a conventional criterion based on the Glauber $\cal P$-function but possessing non-zero Gaussian quantum discord.   Their quantum nature is experimentally revealed by acting locally on one part of the discordant state.  Adding an environmental system purifying the state, we unveil the flow of quantum correlations within a global pure system using the Koashi-Winter inequality.  We experimentally verify and investigate the counterintuitive effect of discord increase under the action of local loss and link it to the entanglement with the environment. For a discordant state generated by splitting a state in which the initial squeezing is destroyed by random displacements, we demonstrate the recovery of entanglement highlighting the role of system-environment correlations.
\end{abstract}
\maketitle

As quantum information science develops towards quantum
information technology, the question of the efficient use and
optimization of resources becomes a burning issue. So far, quantum
information processing (QIP) has been mostly thought of as {\it
entanglement}-enabled technology. Quantum cryptography is an
exception, but even there the so-called effective entanglement
between the parties plays a decisive role  \cite{Haeseler10,
Khan13}.
\setcounter{page}{1}
With the advent of new quantum computation paradigms
\cite{Knill_98} interest in more generic and even non-entangled
QIP resources has emerged \cite{Datta_08}. Unlike
entanglement, the new resources, commonly dubbed as quantum
correlations, reside in all states which do not diagonalize in any
local product basis. While for pure states entanglement and
quantum correlations are equivalent notions, this is not the case
for mixed states. Quantumness of correlations in separable mixed states
is fundamentally related to the non-commutativity of observables, non-orthogonality of states and properties of quantum measurements,
whereas entanglement can be seen as a consequence of the quantum
superposition principle. Correlated mixed states are a lucid illustration of the
fact that the quantum-classical divide  is actually
purpose-oriented and that such states, long considered
unsuitable for QIP, may become a robust and efficient quantum
tool. This may change our understanding of what the ultimate QIP
resources are.

In what follows, we will use quantum discord \cite{discord} for
quantification of quantum correlations. For two systems $A$ and
$B$, quantum discord is defined as the difference,
\begin{eqnarray}\label{discord}
\mathcal{D}^\leftarrow({AB}) &=& {\cal I}({AB})
-\mathcal{J}^\leftarrow({AB}),
\end{eqnarray}
between quantum mutual information ${\cal I}({AB})={\cal
S}({A})+{\cal S}({B})-{\cal S}({AB})$ encompassing all
correlations present in the system, and the one-way classical
correlation $\mathcal{J}^\leftarrow({AB})={\cal
S}({A})-\inf_{\{\hat\Pi_i\}} {\cal H}_{\{\hat\Pi_i\}}(A|B)$, which
is operationally related to the amount of perfect classical
correlations which can be extracted from the system
\cite{Devetak_04}. Here, ${\cal S}$ is the von Neumann entropy of
the respective state, ${\cal H}_{\{\hat\Pi_i\}}(A|B)$ is the
conditional entropy with measurement on $B$, and the infimum is
taken over all possible measurements $\{\hat\Pi_i\}$.

In this paper, we focus on quantum correlations in bi-partite mixed Gaussian states relevant in the context of continuous-variable quantum information \cite{QuInfo}. The
respective correlation quantifier is then Gaussian quantum discord
\cite{adesso,paris} defined by Eq.~(\ref{discord}), where the
minimization in $\mathcal{J}^\leftarrow({AB})$ is restricted to
Gaussian measurements. The Gaussian discord coincides with unrestricted
discord (\ref{discord}) for some states considered by us \cite{Pirandola_14, suppl} which
confirms the relevance of its use. All non-product bi-partite Gaussian states
have been shown to have non-zero Gaussian discord \cite{adesso,Mista_14}
but many of them are termed classical according to the conventional
nonclassicality criterion. That is, their density matrix $\hat \rho$ can be represented as a statistical mixture of two-mode coherent states $|\alpha\rangle|\beta\rangle$  with well behaved $\mathcal{P}$-function,
$\hat{\rho}=\int\int_{\mathbb{C}}\mathcal{P}(\alpha,\beta)|\alpha\rangle\langle\alpha|\otimes|\beta\rangle\langle\beta|d^2\alpha d^2\beta$ \cite{Glauber_63}.
Thus a wide range of states, normally perceived as classical,
exhibit quantum correlations according to the Gaussian discord
and should be classified as quantum. Recurring examples of
non-zero Gaussian discord in such seemingly classical states
raised doubts whether Gaussian discord is a legitimate measure.
This apparent discrepancy was discussed in \cite{ferraro-paris}:
the nonclassicality criteria can differ in the quantum-optical
realm and in information theory. Therefore states classified as
quantum in one context, can appear classical in the other. We
provide experimental and theoretical evidence that the quantum
nature of the bi-partite mixed separable states is correctly
captured by non-zero Gaussian discord and this quantumness can be
revealed by acting merely locally on one part of the state.

Gaussian states are quantum states of systems in
infinitely-dimensional Hilbert space, e.~g. light modes, which
possess a Gaussian-shaped Wigner function. Correlations carried by a Gaussian state of two modes $A$ and
$B$ are thus completely characterized by the covariance matrix
(CM) $\gamma$ \cite{cov-matrix} with elements
$\gamma_{ij}=\langle\hat{\xi}_i \hat{\xi}_j+\hat{\xi}_j
\hat{\xi}_i\rangle-2\langle\hat{\xi}_i\rangle\langle\hat{\xi}_j\rangle$,
where
$\mathbf{\hat{\xi}}=(\hat{x}_A,\hat{p}_A,\hat{x}_B,\hat{p}_B)$ is
the vector of quadratures. A Gaussian state with CM $\gamma$ is

separable iff $\gamma^{(T_{A})}+i\Omega\geq0$ \cite{Simon_00},
where $\gamma^{(T_{A})}=L\gamma L^{T}$ with $L=\mbox{diag}(1,-1,1,1)$ and
$\Omega=\oplus_{j=1}^{2}i\sigma_{y}$, where $\sigma_{y}$ is the
Pauli-$y$ matrix. Gaussian discord carried by the state can be determined from $\gamma$ using the analytic formula
derived in \cite{paris,suppl}.

\begin{figure}[t]
\begin{center}
\includegraphics[width=8.0cm]{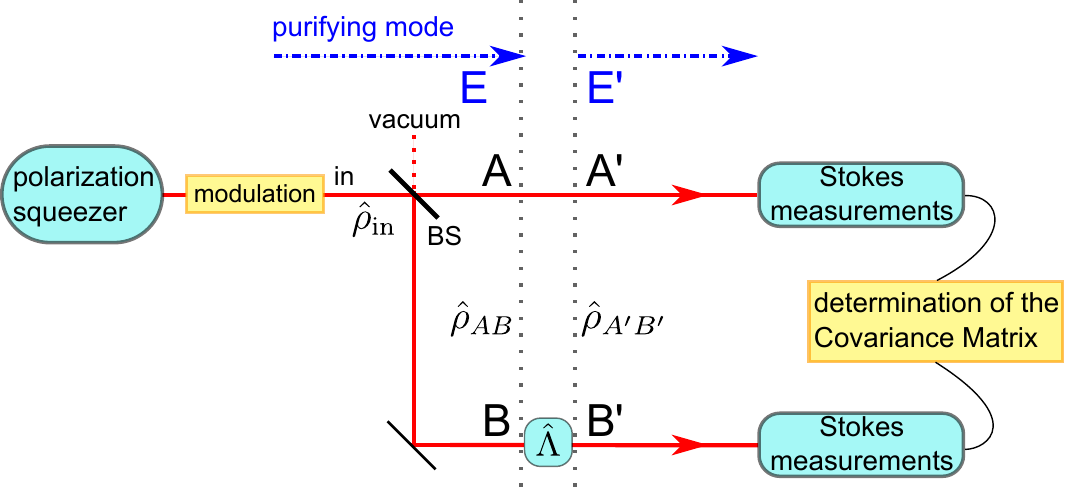}
\end{center}
\caption{Experimental scheme. BS: beamsplitter,
$\hat{\rho}_{\rm in}$: input state prepared by Gaussian
distributed modulation of coherent or squeezed states, $E$
($E^\prime$): environmental mode purifying mixed state
$\hat{\rho}_{AB}$ ($\hat{\rho}_{A'B'}$), $\hat{\Lambda}$: local
operation on $B$. }\label{scheme}
\end{figure}

\noindent \textit{Discord increase
under local loss.} We prepare a coherent or squeezed optical mode and add
noise in the form of Gaussian-distributed random displacements of
the $x$-quadrature. The optical mode is then in a classical
state given by a convex mixture of coherent states and is
split up on a beamsplitter as depicted in Fig.~\ref{scheme}. The
output two-mode state after the BS with CM $\gamma_{AB}^{\rm coh}$ ($\gamma_{AB}^{\rm sq}$) has a non-zero
Gaussian discord despite being classical according to the $\cal P$-function criterion.
These quantum states exhibit notable
robustness against noise and coupling to environment. Indeed, as was first shown theoretically  for qubits \cite{streltsov,giovannetti,campbell}, quantum correlations can even emerge from a purely classically correlated state under the action of a local noise.  This work was then
extended to Gaussian states \cite{giovannettiCV}, and
discord increase under local loss has been experimentally demonstrated
\cite{andersen}.

The experimental setup is shown in Fig.~\ref{scheme}. The coherent mode utilized in one of the experiments described here stems directly from a femtosecond laser. The squeezed state used in the other experiment is implemented as a polarization squeezed beam generated by exploiting the non-linear Kerr effect of a polarization maintaining fiber~\cite{Korolkova2002, Heersink_05, Dong_07}. For practical reasons the quantum states are encoded in polarization variables and measured by a Stokes detection. Using intense light fields the Stokes observable in the dark plane $\hat S_\theta$ is associated with $\hat x$ and $\hat S_{\theta+\pi/2}$ with $\hat p$~\cite{Heersink_05}. The squeezed Stokes observable is modulated by an electro-optical modulator (EOM) at the sideband frequency of 18.2\,MHz. This is equivalent to a displacement in the dark plane of the quantum states defined at this sideband frequency. Then the mode is divided on a balanced beamsplitter (BS). The modes are detected by Stokes measurements and the signal is down-converted at the modulated sideband frequency. The data taken for different displacements is combined computationally to prepare a Gaussian mixed two-mode state. The modulation patterns are chosen such that the initial squeezing is destroyed and the state $\hat{\rho}_{AB}$ is separable. The Stokes measurements allow the determination of its complete CM.

\begin{figure}[t]
\includegraphics[width=7cm]{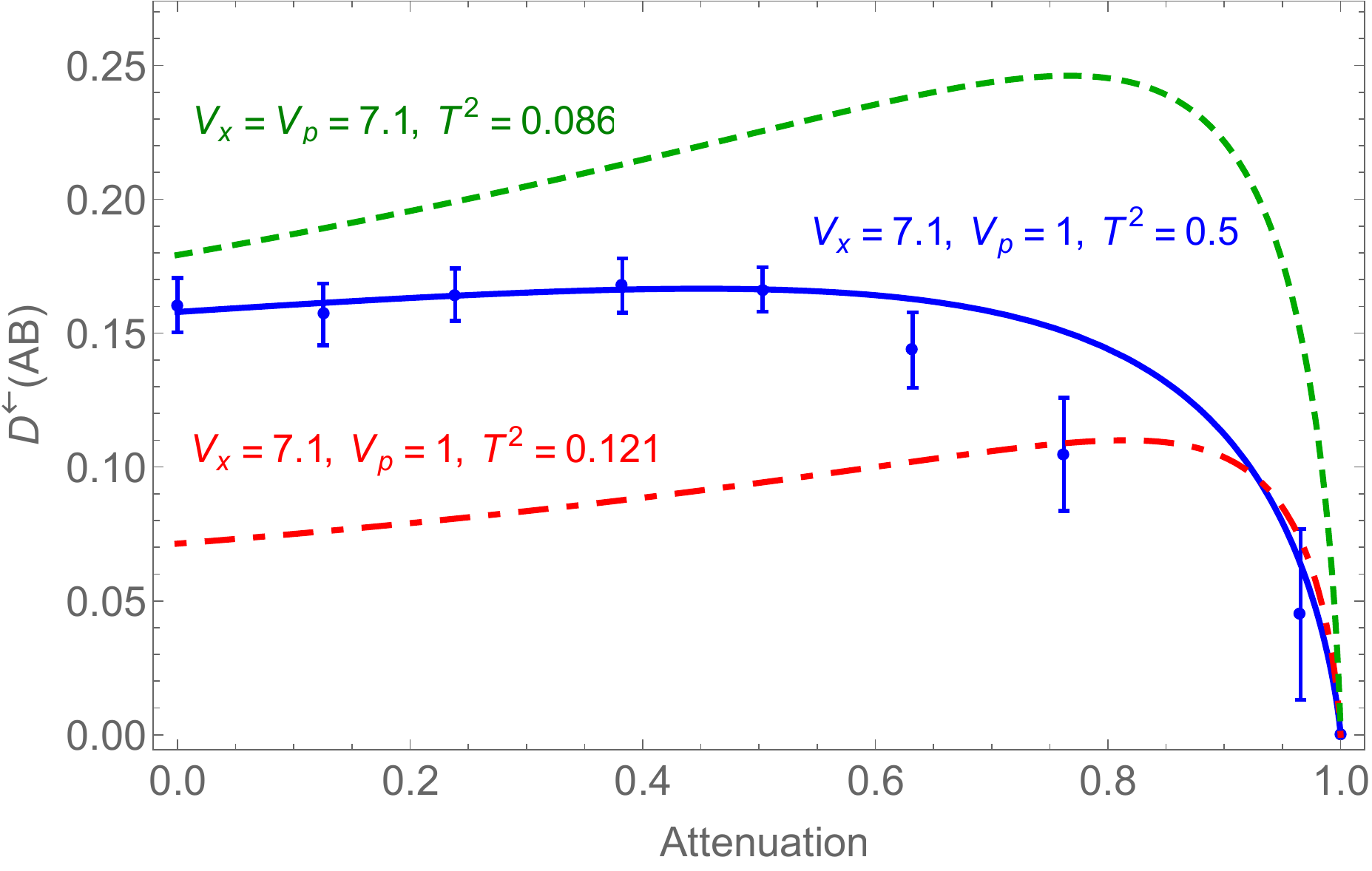}
\caption{ (color online). Quantum discord $\mathcal{D}^\leftarrow(\hat{\rho}_{AB})$ versus
attenuation in mode $B$ for modulated coherent state. Theory curve
(solid blue) and experiment (blue dots) for modulation in
$x$-quadrature, $V_x=7.1$, $V_p=1$ and $T^2=0.5$. Theory curves
for the same input state and $T^2=0.121$ (red dot-dashed) and for modulation
in both quadratures, $V_x=V_p=7.1$, and $T^2=0.086$ (green dashed). }
\label{discord-coh}
\end{figure}

Since quantum discord is related to the
non-commutativity of observables, it is often expected that
modulation in both conjugate quadratures is required to see
quantum behaviour. In contrast to all previous
discord experiments \cite{andersen,vedral,vogl}, to generate
discord we modulate the input coherent state only in one of the
conjugate quadratures, $\hat x_{\rm in}$, keeping $\hat p_{\rm in}$ at the
coherent-state level, $V_{p}=2\langle \hat p^2_{\rm in} \rangle =1$, where
$\hat{x}_{\rm in}$ ($\hat{p}_{\rm in}$) is the $x$($p$)-quadrature
of mode
``in''.
The local loss
is realized by variable attenuation in mode $B$ denoted as
$\hat{\Lambda}$ in Fig.~\ref{scheme}.  The highest
discord in $\gamma_{A'B'}^{\rm coh}$
is achieved when the initial mode is split on a balanced BS. Up to a certain attenuation level $\mathcal{D}^\leftarrow(\hat{\rho}_{AB})$  grows monotonically with
increasing modulation depth, i.e., with $V_x=2\langle\hat{x}_{\rm in}^2\rangle$, and finally drops sharply. Gaussian states with CM $\gamma_{AB}^{\rm coh}$ are convex mixtures of non-orthogonal overcomplete coherent basis states. The impossibility to deterministically discriminate between non-orthogonal states is a seminal example of quantumness in separable bipartite states.  Intuitively the discord growth under the action of local loss can be attributed to these non-orthogonal basis states becoming less distinguishable with attenuation, although, as previous work shows \cite{streltsov,giovannetti,campbell}, it is difficult to reduce the mechanism behind this effect to a simple single phenomenon.
The discord rises only very slowly
with loss (Fig.~\ref{discord-coh}, blue dots and solid) as, in addition to its positive role, attenuation renders the CM
$\gamma_{A'B'}^{\rm coh}$  increasingly asymmetric regarding
$A'$ and $B'$
which supresses the discord growth.  The gradient in discord can be
substantially increased by
using an asymmetric BS
when splitting the ``in" mode, such that most of the input beam is reflected into the attenuated mode $B$ (Fig.~\ref{discord-coh}, red dot-dashed).
One can obtain the same effect  by using the balanced
BS and adding asymmetric noise to the CM $\gamma_{A'B'}^{\rm coh}$, which
reflects a limited balancing of the homodyne detectors \cite{andersen}.

Although the quantum effects are observed already when a single quadrature is modulated, displacement in both non-commuting variables
does play an important role. Fig.~\ref{discord-coh} (green dashed) shows the discord increase with attenuation for the input state
equally modulated in both quadratures and for the asymmetric BS. There is an obvious advantage in value and gradient of discord.
Incidentally, these dynamics correspond to the measurement results presented in \cite{andersen}, where the additional ``noise" (imitating scenario with an asymmetric BS) stems from the imperfections in the detection system.

To get a good agreement of theory and experiment (blue dots and solid in Fig.~\ref{discord-coh}) we had to include imperfect common mode rejection (CMR) in homodyne
detection into our theory model. Similar to \cite{andersen} we model
the imperfection by addition of an uncorrelated noise in modes $A'$ and $B'$
which decreases linearly with attenuation in mode $B'$. An even better fit can be achieved without the additional noise, only by using a highly unbalanced BS.

\begin{center}
\begin{figure}[t]
\includegraphics[width=7cm]{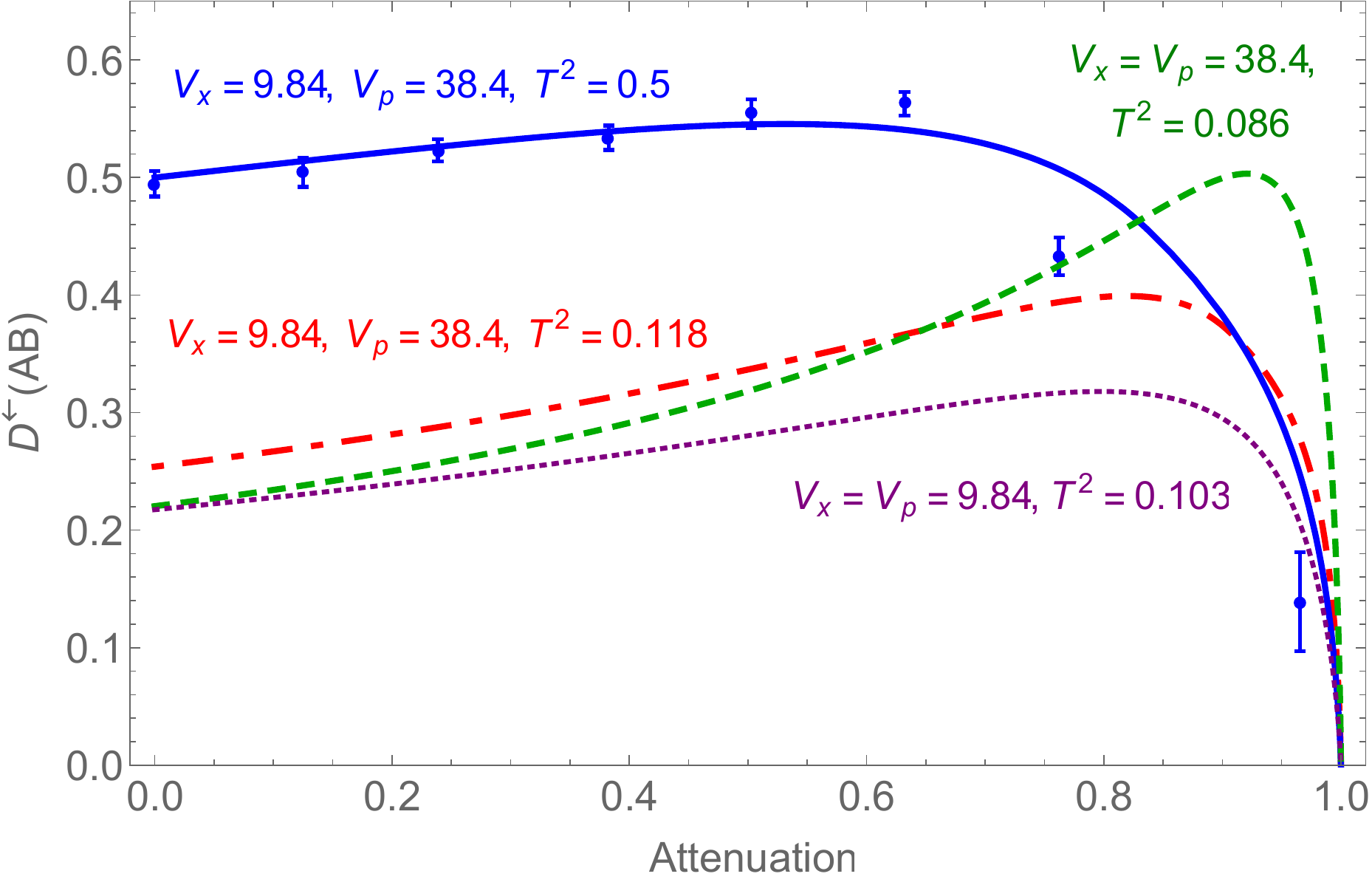}
\caption{ (color online). Quantum discord versus
attenuation in mode $B$ for modulated squeezed state. Theory curve
(blue solid) and experiment (blue dots) for modulation in
$x$-quadrature, $V_x=9.84$, $V_p=38.4$ and $T^2=0.5$. Theory
curves: for the same input and  $T^2=0.118$ (red dot-dashed);  for $V_x=V_p=38.4$
and $T^2=0.086$ (green dashed);  for
$V_x=V_{p}=9.84$ and  $T^2=0.103$ (purple dotted).}
\label{discord-sq}
\end{figure}
\end{center}

There are several important messages here. First, the largest effect of quantum discord increase under local loss is obtained when the output state is symmetrized with respect to quantum uncertainties in modes $A'$ and $B'$. In our case this is achieved by using the asymmetric BS, with the optimal ratio determined by the form of the ``in" CM (cf. red dot-dashed and blue solid theory curves, Fig.~\ref{discord-coh}). Notably,
losses can be turned into positive control mechanism when
using discord as a resource. For example, in case of imperfect CMR modelled by the asymmetric BS,
initial discord is lower (red dot-dashed), which, however, can be
counteracted by including attenuation in $B$ so that this
additional loss closes the gap between the discord values for the asymmetric and symmetric BS. The effect is even more pronounced for the initial state
symmetrically modulated in both quadratures (green dashed), and enhances further when the modulation in both quadratures gets higher (Fig.~\ref{discord-sq}, cf. green-dashed and blue solid curves).
Finally, modulation in both incompatible observables is
advantageous but not always a prerequisite. As a future work, more rigorous analysis
is required to identify the role of both conjugate variables.

It is interesting to explore whether using a quantum resource
initially can bring an advantage.
In contrast
to \cite{andersen}, the input mixed state in Fig.~\ref{discord-sq}
is created by displacing a squeezed state with approximately $-3$
dB squeezing. Although we still displace
the state only along the $x$ axis, it is naturally
blurred also in $p$-quadrature due to the anti-squeezing and the
additional phase noise coming from the propagation in the fiber.
This gives an extremely large $p$-quadrature variance,
$V_p=38.4$. For the discord increase, the only
advantage is through these large input variances, irrespective
of the quantumness initially present (Fig.~\ref{discord-sq}). However, this initial
quantumness does carry a potential to enrich the resultant
discordant state. For example, entanglement which would emerge
after the BS if no displacement is performed, can still be
recovered and used, as we show in the last section.

System-environment correlations provide another control mechanism when using correlated mixed states and give a deeper insight into the quantum effects related to non-zero discord. Assume, there is a
third mode $E$ carrying maximum information about the state
$\hat{\rho}_{AB}$, that might be imprinted onto the environment (Fig.~\ref{scheme}).
The global state of the system is then the purification $|\psi
\rangle_{ABE}$ of $\hat{\rho}_{AB}$,
$\mbox{Tr}_{E}(|\psi
\rangle_{ABE}\langle\psi|)=\hat{\rho}_{AB}$. The initial
purification before the BS is a locally squeezed two-mode squeezed
vacuum state $|\psi\rangle_{AE}$. Note that the purification for
any discordant state is entangled across the $E-(AB)$ splitting, which already links discord and entanglement with the environment. To analyze further the flow of correlations in a global system $|\psi \rangle_{ABE}$, we
apply the Koashi-Winter relation \cite{koashi}
\begin{eqnarray}
\label{Koashi}
\mathcal{S}\left(A\right)=\mathcal{E}_F\left({AE}\right)+\mathcal{J}^\leftarrow\left({AB}\right),
\end{eqnarray}
which connects the marginal entropy $\mathcal{S}\left(A\right)$,
one-way classical correlation
$\mathcal{J}^\leftarrow\left({AB}\right)$ and entanglement of
formation (EoF) $\mathcal{E}_F\left({AE}\right)$.
The classical correlation
$\mathcal{J}^\leftarrow\left({AB}\right)$ is directly linked to
discord (see Eq.~\ref{discord}).

In our scheme (Fig.~\ref{scheme}), both mutual information and classical correlation in the discord definition decrease with
attenuation, but at different rates resulting in an overall
discord increase. As the marginal entropy of $A$ remains unchanged
under attenuation in mode $B$, for the relation,
$\mathcal{S}\left(A\right)=\mathcal{E}_F\left({A^\prime
E^\prime}\right)+\mathcal{J}^\leftarrow\left({A^\prime
B^\prime}\right)$ to hold the decrease in classical correlation
$\mathcal{J}^\leftarrow\left({A^\prime B^\prime}\right)$ has to be
accompanied by increase in $\mathcal{E}_F\left({A^\prime
E^\prime}\right)$ between the unmeasured mode $A'$ and the
environment.

\begin{center}
\begin{figure}
\includegraphics[width=7.0cm]{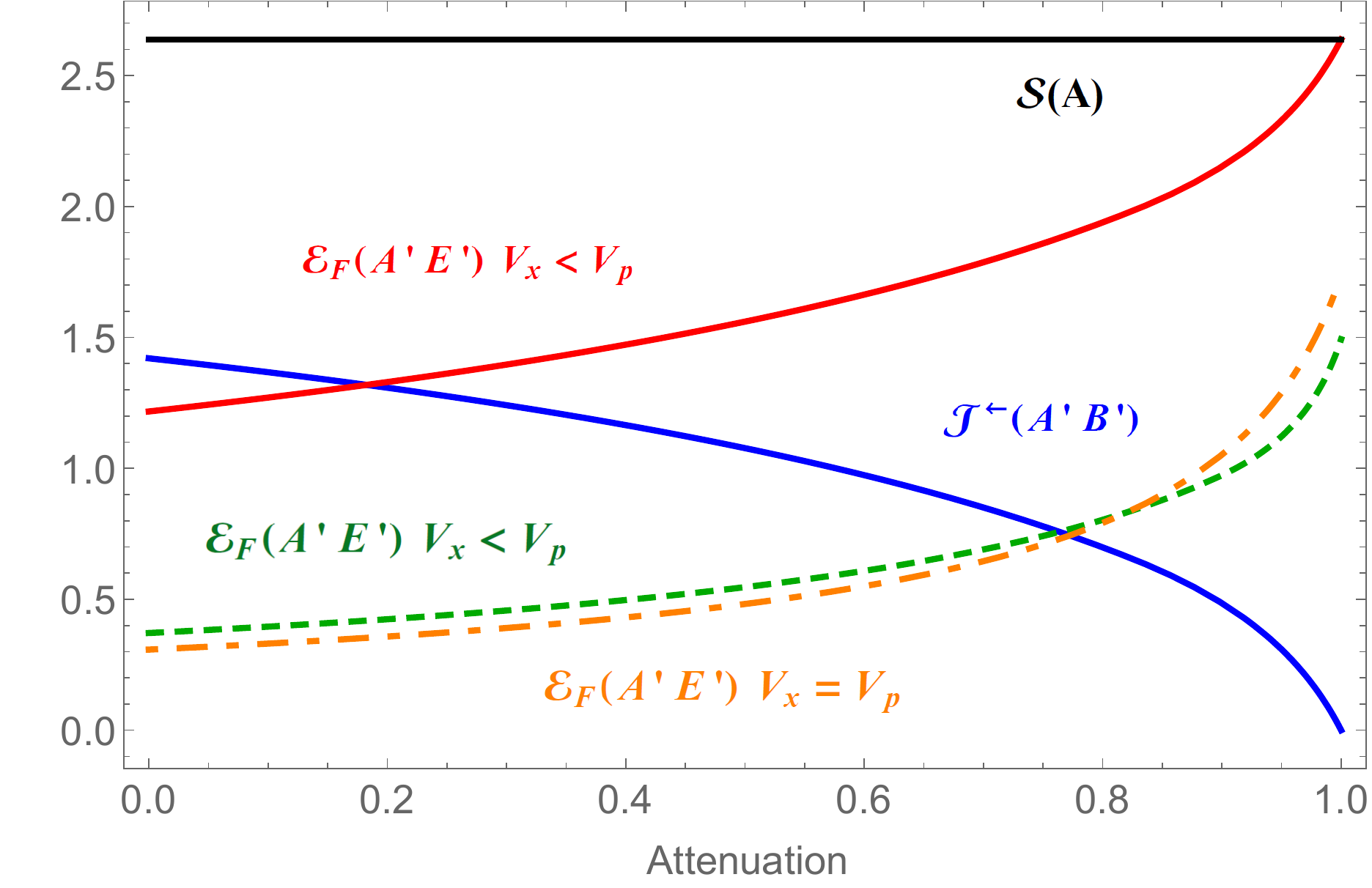}
\caption{Flow of quantum correlations in the global
state $|\psi\rangle_{ABE}$ for the states shown in Fig.~\ref{discord-sq}.
Classical correlation (solid blue curve), marginal
entropy (solid black line), system-environment EoF for $V_x=9.84,
V_p=38.4$ and $T^2=0.5$ (solid red curve).  System-environment EoF for same input state and $T^2=0.118$ (green
dashed); for the symmetric input state $V_{x}=V_p=38.4$ and $T^2=0.086$ (orange
dot-dashed).} \label{koashi-graph}
\end{figure}
\end{center}

The results for the flow of correlations are presented in
Fig.~\ref{koashi-graph} for the input states used in
Fig.~\ref{discord-sq}. For computing the EoF of a general Gaussian
state $\hat \rho_{AE}$ we used the technique of
Ref.~\cite{adesso-EoF}. As clearly seen, the growth of discord
relates to the increasing EoF with the environment. Fig.~\ref{koashi-graph}
also witnesses that the Koashi-Winter relation holds for this type of Gaussian states. For the
experimentally measured case of Fig.~\ref{discord-sq}, the rising
entanglement with environment and decreasing classical correlation
between the system modes $A$ and $B$ add up to the constant
marginal entropy $\mathcal{S}\left(A\right)$. We have also
verified that if a measurement is performed on mode $A$, discord
always decreases, as does entanglement with environment
$\mathcal{E}_F\left(A^\prime E^\prime\right)$.  In the qubit case,
the role of system-environment correlations is particularly
eloquent and increase in discord in both cases can be enacted by
performing the entangling operation on $A$ and some environmental
mode, instead of locally attenuating $B$ \cite{tatham}.  Recently,
a further experiment has been proposed linking the open-system
dynamics of entanglement to correlations with environment and
discord \cite{aguilar}.

\noindent \textit{Entanglement recovery.} Consider now the state
$\hat{\rho}_{AB}$ prepared from a state with $-3$ dB of squeezing in $x$-quadrature using
Gaussian modulation in the same quadrature. The measured CM reads
\begin{align}
\gamma_{AB}^{\rm sq}= \left(
\begin{array}{cccc}
  5.42 & 0.23 & 4.06 & 0.04 \\
  0.23 & 19.28 & 0.45 & 17.29 \\
  4.06  & 0.45  & 4.73 & 0.55 \\
 0.04 & 17.29 & 0.55 & 17.70 \\
 \end{array}\right), \label{CM-ABsq}
\end{align}

where the measurement errors are given in \cite{suppl}.
The local CMs are not squeezed which verifies
that the displacements in the direction of squeezing destroyed the
squeezing \cite{global_squeezing, sq_explanation}. The state of modes $A$ and $B$ is then
inevitably separable \cite{Kim_02} as witnessed by the
nonnegativity of the minimal
eigenvalue $\mbox{min}\{\mbox{eig}[(\gamma_{AB}^{\rm
sq})^{(T_{A})}+i\Omega]\}=0.84\pm0.02$.
However the state contains quantum correlations as evidenced by
$\mathcal{D}^\leftarrow({AB})=0.49\pm0.01$. The correlations originate from two
sources.
First, the random displacement $\bar{x}$ of the
$x$-quadrature of the input mode ``in'' yields quantum
correlations between separable modes $A$ and $B$ exactly as in the
case of coherent initial state. Secondly, the initial
squeezing of mode ``in'' would alone create entanglement
between $A$ and $B$.

Interestingly, there exists a scenario, in which
correlations of the system $(AB)$ with a separable environmental
mode $\tilde E$ allow to eliminate the displacement noise and
recover this entanglement between  $A$ and $B$. Note, that mode $\tilde E$ is not purifying.
Preparation of the state with CM (\ref{CM-ABsq}) by splitting a
randomly displaced squeezed input mode ``in" on BS is in fact the
preparation of the two-mode reduced state in the entanglement sharing
protocol \cite{Mista_13}. Imagine that like in the protocol, the
$x$-quadrature of $\tilde{E}$ encodes
the random displacement $\bar{x}$ as $x_{\tilde E}-\bar{x}$. In
contrast to the previously considered purifying mode $E$, mode $\tilde{E}$ has
been created by local operations and classical communication (LOCC)
and hence it is separable from the subsystem.
Next, as in \cite{Mista_13},  mode $B$ is transmitted to the
location of mode $\tilde{E}$ where the modes are superimposed on a
beamsplitter BS$_{B\tilde{E}}$. As a consequence, the noise caused by the random
displacements is partially cancelled and the entanglement between
modes $A$ and $B$ is restored as desired.

This entanglement recovery reveals two important facts about quantum correlations in the global system $(AB\tilde{E})$.
First, it demonstrates that there must exist entanglement
across the $A-(B\tilde{E})$ splitting before the beamsplitter BS$_{B\tilde{E}}$ as otherwise
it could not create entanglement between modes $A$ and $B$. Second, it is a proof that mode $B$ shares quantum correlations with
the subsytem $(A\tilde{E})$ and therefore realizes a true quantum
communication between the locations of modes $A$ and $\tilde{E}$, which cannot
be replaced by LOCC. Indeed, if mode $B$ was only classically correlated
with subsystem $(A\tilde{E})$, it would be possible to replace
its transmission by a measurement of its state (which does not disturb the
global state), followed by recreation of the state in the location of mode
$\tilde{E}$. This is, however, an LOCC operation which cannot establish
entanglement across $A-(B\tilde{E})$ splitting.

Instead of physically imprinting a displacement on the third quantum
mode $\tilde{E}$ and interfering the mode with mode $B$ on a beamsplitter,
we have superimposed mode $B$ with vacuum mode $\tilde{E}$ on a beamsplitter and implemented equivalent displacement electronically on the
measured data. This gives us a violation of
Duan's separability criterion~\cite{Duan_00,1,2,suppl} $0.91\pm0.01<1$, which certifies entanglement between $A$ and $B$.

If we have access to the displacement $\bar{x}$ encoded on mode $\tilde{E}$,
entanglement between modes $A$ and $B$ can be recovered by directly
performing the reverse displacement on mode $B$ to cancel the modulation.
By executing this computationally, we got a violation of
Duan's separability criterion
of $0.83\pm0.01 < 1$ proving that modes $A'$ and $B'$ after demodulation are entangled (see \cite{suppl} for details).
The fact that the entanglement can also be recovered by imprinting the available information about the state preparation directly locally on mode $B$ features the important role
the communication of classical information can play in quantum communication, in particular when
using separable discordant states
(cf.~\cite{peuntinger13}).

In summary, we have demonstrated the role and utility of system-environment correlations in discord dynamics
and provided new insights into discord increase under dissipation and quantumness of these correlations.

L. M. acknowledges project P205/12/0694 of GA\v{C}R.
N. K. is grateful for the support provided by the A. von Humboldt Foundation. N. Q. and N. K. acknowledge the support from the Scottish Universities Physics Alliance (SUPA) and the
Engineering and Physical Sciences Research Council (EPSRC).
The project was supported within the framework of the BMBF grant ``QuORep'' and in the framework of the International Max Planck Partnership (IMPP) with Scottish Universities.\\

\begin{appendix}

\section{Quantumness of Gaussian Discord\\
Supplementary Information}

\subsection{Gaussian Quantum Discord}

\subsubsection{Standard Form Covariance Matrix}
Using local symplectic transformations, every covariance matrix (CM) describing a bipartite state can be expressed in standard form \cite{Simon_00_sup}:
\begin{equation}\label{standardform}
\gamma=
\begin{pmatrix}
\alpha&\delta\\
\delta^T&\beta
\end{pmatrix}
=
\begin{pmatrix}
a&0&c_+&0\\
0&a&0&c_-&\\
c_+&0&b&0\\
0&c_-&0&b
\end{pmatrix}.
\end{equation}
From this standard form we may define local invariants such as the seralian, defined as the sum of the determinants of the four subblocks $\Delta=a^2+b^2+2c_+c_-$. The seralian in turn can be used to define the symplectic eigenvalues \cite{Williamson_36_sup} of the CM as
\begin{equation}
\nu_{\pm}=\sqrt{\frac{\Delta \pm \sqrt{\Delta^2-4\text{det}\gamma}}{2}}.
\end{equation}


\subsubsection{Definition}

The Gaussian quantum discord using Gaussian measurements and von Neumann entropies was put forward by Adesso {\it{et al.}} in \cite{adesso_sup}, stating that
\begin{equation}\label{D}
\mathcal{D}^\leftarrow_{AB}=f\left(\sqrt{B}\right) - f(\nu_-) - f(\nu_+) + \inf_{\sigma_0} f\left(\sqrt{\det\epsilon}\right)\\
\end{equation}
where $\nu_{\pm}$ are the symplectic eigenvalues of the two-mode CM for modes A and B and
$
f(x)=\left(\frac{x+1}{2}\right)\ln\left(\frac{x+1}{2}\right)-\left(\frac{x-1}{2}\right)\ln\left(\frac{x-1}{2}\right).
$
The optimal determinant of the CM
$\varepsilon$ of the post-measurement state,
$\mbox{inf}\mbox{det}\varepsilon$, is given by
\begin{equation}
\inf_{\sigma_0}\det\epsilon=\begin{cases}
\frac{2 C^2 + \left(B - 1\right) \left(D - A\right) + 2 \left\vert C\right\vert\sqrt{C^2 + \left(B - 1\right) \left(D - A\right)}}{\left(B - 1\right)^2}\\ \text{if}\quad\left(D - A B\right)^2 \leq \left(1 + B\right) C^2 \left(A + D\right),\\
\frac{A B - C^2 + D - \sqrt{C^4 + \left(D - A B\right)^2 - 2 C^2 \left(A B + D\right)}}{2 B}\\ \text{Otherwise}.
\end{cases}
\end{equation}
where $A=\det \alpha$, $B=\det \beta$, $C=\det \delta$ and $D=\det \gamma$. The
term $\mbox{inf}_{\sigma_0}f(\sqrt{\mbox{det}\varepsilon})$ represents
optimized average of von Neumann entropies of states of mode $A$
conditioned on the outcomes of a Gaussian measurement with CM $\sigma_{0}$
on mode $B$. The separate scenarios refer to different types of Gaussian measurements, a notable class of states falling under the first case are squeezed thermal states for which the conditional measurement is minimized by heterodyne measurements, the second case corresponds to homodyne measurements. Note that the directionality of the arrows in the above indicates on which subsystem a measurement has been performed, in the present case a measurement is performed on subsystem $B$.

Quantum discord quantifies quantum correlations in a quantum state
$\rho_{AB}$ as an amount of information about a quantum system $A$
which cannot be extracted by performing the best measurement on system
$B$ \cite{discord_sup}, which gains the maximum information about $A$. In this
paper, we quantify quantum correlations by means of the Gaussian
discord \cite{discord_sup} for which the best measurement is always picked
from the set of Gaussian measurements. One may argue then, that a
more relevant quantifier of quantum correlations would be the
more general quantum discord admitting also non-Gaussian
measurements, which can in principle extract strictly more
information than any Gaussian measurement. Needless to say,
this can really be the case for some two-mode Gaussian states,
including states obtained by splitting a modulated coherent
state on a beam splitter, which we consider here. Nevertheless, a recent result of
Ref.~\cite{Pirandola_14_sup} reveals, that for the other class of states considered
here, which are prepared by splitting a modulated squeezed state,
out of all possible measurements (including non-Gaussian ones) the best
measurement is always Gaussian. This implies that Gaussian discord
coincides with discord for these states, which speaks in favor of
the present use of the discord (\ref{D}) as a relevant quantifier
of quantum correlations.

Moving to the proof of the optimality of Gaussian measurements for
the latter states, we start by writing down their CM as
\begin{equation}\label{gam}
\gamma_{AB}=\frac{1}{2}
\begin{pmatrix}
\gamma_{\rm in}+\openone&\gamma_{\rm in}-\openone\\
\gamma_{\rm in}-\openone&\gamma_{\rm in}+\openone\\
\end{pmatrix},
\end{equation}
where $\gamma_{\rm in}=\mbox{diag}(V_{x},V_{p})$ and $\openone$ is
the $2\times2$ identity matrix. We assume that the state with CM
$\gamma_{\rm in}$ has been prepared by Gaussian distributed random
displacement of the quadrature $x$ of a squeezed state with
squeezing in the quadrature $x$ and large antisqueezing in
quadrature $p$, and hence $V_{p}>V_{x}>1$. By means of local
squeezing transformations we can bring the CM (\ref{gam}) to the
standard form (\ref{standardform}) with
\begin{align}\label{abcpmour}
&a=b=\frac{\sqrt{(V_{x}+1)(V_{p}+1)}}{2},\quad
c_{+}=\sqrt{\frac{V_p+1}{V_{x}+1}}\left(\frac{V_{x}-1}{2}\right),\quad \nonumber\\
&c_{-}=\sqrt{\frac{V_x+1}{V_{p}+1}}\left(\frac{V_{p}-1}{2}\right).
\end{align}
In Ref.~\cite{Pirandola_14_sup} the optimality of Gaussian measurement
was proven for all two-mode Gaussian states $\rho_{AB}$, which can
be decomposed as
\begin{equation}\label{rhoAB}
\rho_{AB}=[\mathscr{S}_{A}(\xi)\mathcal{E}_{A}\mathscr{S}_{A}^{-1}(r)\otimes\mathcal{I}_{B}](\rho_{AB}^{TMSV}).
\end{equation}
Here, $\rho_{AB}^{TMSV}$ is the two-mode squeezed vacuum state
with CM
\begin{equation}\label{TMSV}
\gamma_{AB}^{TMSV}=
\begin{pmatrix}
m\openone& \sqrt{m^2-1}\sigma_{z}\\
\sqrt{m^2-1}\sigma_{z} & m\openone\\
\end{pmatrix}
\end{equation}
and $\mathcal{E}$ is a single-mode Gaussian channel, which acts on
a single-mode CM $\Gamma$ as $\Gamma'=X\delta X^{T}+Y$, where
$X=\sqrt{|\tau|}\mbox{diag}[1,\mbox{sign}(\tau)]$ and
$Y=\eta\openone$. Further, $\mathscr{S}(r)$ and $\mathscr{S}(\xi)$
are single-mode squeezing operations, which are represented at the
level of CM by the diagonal matrix
$S(t)=\mbox{diag}(t^{\frac{1}{2}},t^{-\frac{1}{2}})$, $t=r,\xi$,
where
\begin{equation}\label{xi}
\xi=r\frac{\theta{(r^{-1})}}{\theta(r)}, \quad
\theta(r)\equiv\sqrt{\eta r+|\tau|m},
\end{equation}
is chosen such that the state (\ref{rhoAB}) has CM in the standard
form (\ref{standardform}). Finally, the parameters $\tau$, $\eta$, $r$ and $m$ must satisfy
conditions
\begin{equation}\label{conditions}
\tau\in\mathbb{R},\quad \quad\eta\geq|1-\tau|,\quad
r\in[m^{-1},m].
\end{equation}

By expressing the CM of the state (\ref{rhoAB}) using
Eqs.~(\ref{TMSV}) and (\ref{xi}) we get the elements of the
standard-form CM (\ref{standardform}) for states for which the
Gaussian discord is optimal \cite{Pirandola_14_sup},
\begin{align}\label{abcpm}
&a=\theta(r)\theta(r^{-1}),\quad
c_{+}=\sqrt{|\tau|(m^2-1)\frac{\theta(r^{-1})}{\theta(r)}},\quad \nonumber\\
&c_{-}=-\mbox{sign}(\tau)\sqrt{|\tau|(m^2-1)\frac{\theta(r)}{\theta(r^{-1})}}.
\end{align}
Let us now show that under certain condition on $V_{x}$ which is
satisfied in the present case, one can really express the
standard-form elements (\ref{abcpmour}) as in Eq.~(\ref{abcpm}) while
fulfilling conditions (\ref{conditions}). Indeed, by equating
right-hand sides of Eqs.~(\ref{abcpmour}) and (\ref{abcpm}) and
expressing $m,\tau,\eta$ and $r$ via $V_{x}$ and $V_{p}$, one
finds after some algebra that $m=a$,
\begin{align}\label{rtaum}
&\tau=-\frac{(V_{x}-1)(V_{p}-1)}{(V_{x}+1)(V_{p}+1)-4},\quad
\eta=\frac{2(V_{x}V_{p}-1)}{(V_{x}+1)(V_{p}+1)-4},\quad\nonumber\\
&r=\sqrt{\frac{V_{x}+1}{V_{p}+1}}\left(\frac{V_{p}-1}{V_{x}-1}\right).
\end{align}
First, $\tau$ is real and therefore the first of conditions
(\ref{conditions}) is satisfied. Second, because $\eta=1-\tau$ the
second condition in (\ref{conditions}) is fulfilled and the
channel $\mathcal{E}_{A}$ in the decomposition
(\ref{rhoAB}) is the phase-conjugating channel which can be realized
by the two-mode squeezer where we take idler mode as an output. Third,
as we can write $r=m^{-1}(V_{x}+1)(V_{p}-1)/[2(V_{x}-1)]$, one gets immediately
using inequalities $V_{x}>1$ and $V_{p}>V_{x}$ that $r>m^{-1}$.
Finally, we also have $r=2m(V_{p}-1)/[(V_{x}-1)(V_{p}+1)]$ which gives $r\leq m$ provided that
$3-4/(V_{p}+1)\leq V_{x}$. The left-hand side of the latter inequality is a monotonically increasing
function of $V_{p}$ which approaches the maximum value of $3$ in the limit of infinitely large $V_{p}$. Therefore, for
states with a sufficiently large modulation in the quadrature $x$ such that $V_{x}\geq 3$ also the third condition
(\ref{conditions}) is fulfilled. In the present paper we consider strongly modulated squeezed states which reliably
satisfy the latter inequality as can be easily seen by inspection of the CM in Eq.~(3) of the main paper, which
concludes our proof of optimality of Gaussian discord.

\subsection{Experimental Setup}
\begin{figure}[t]
\begin{center}
\includegraphics[width=1.0\columnwidth]{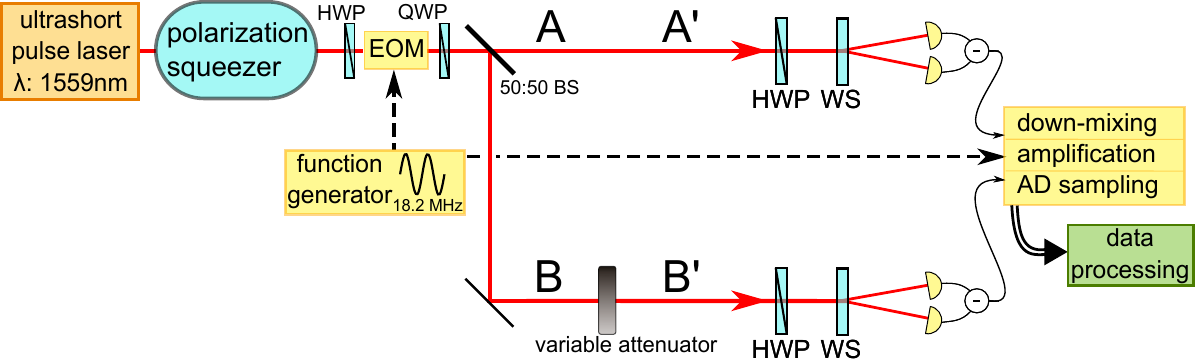}
\end{center}
\caption{Experimental setup. HWP: half-wave plate, EOM: electro-optical
modulator, QWP: quarter-wave plate, BS: beamsplitter, WS: Wollaston prism
}\label{setup}
\end{figure}

In Fig.\,\ref{setup}, the experimental setup is depicted. The description of the general ideas and functionality of the experiment can be found in the main text of the paper. Here, we want to give additional information about the details of the practical implementation.\newline
We use a soliton laser with a pulse length of $\sim$200\,fs at a center wavelength of 1559\,nm (repetition rate: 80\,MHz). For the preparation of the squeezed states, the non-linear Kerr effect of a polarization maintaining fiber (FS-PM-7811, Thorlabs, 13\,m) is exploited to generate polarization squeezing~\cite{Heersink_05_sup, Leuchs99_sup, Silberhorn01_sup, Dong_07_sup}. The squeezed Stokes observable is modulated by an electro-optical modulator (EOM). The applied sinusoidal voltage $V_{\text{mod}}$ generates a sideband at 18.2\,MHz. The modulated Stokes observable $\hat{S}_\theta$ has to be adjusted by a half-wave plate in front of the EOM. To compensate for the stationary birefringence of the EOM, we use a quarter-wave plate.\newline
The such prepared mode is then divided on a 50:50 beamsplitter into the two modes $A$ and $B$. The mode $B$ undergoes a variable attenuation. We then measure the Stokes observables of the two modes by means of the two measurement setups consisting of a rotatable half-wave plate, a Wollaston prism and the difference signal of a pair of PIN photodiode detectors. These Stokes measurements are used to determine the complete covariance matrix by measuring all possible combinations of the squeezed and antisqueezed Stokes variables. The determination of the correlations of squeezed and antisqueezed Stokes observable within one mode is carried out by means of a measurement of the linear combination of these. We thus measure five pairs of observables: $(\hat{S}_{A^\prime,0^\circ}, \hat{S}_{B^\prime,0^\circ})$, $(\hat{S}_{A^\prime,90^\circ}, \hat{S}_{B^\prime,0^\circ})$, $(\hat{S}_{A^\prime,0^\circ}, \hat{S}_{B^\prime,90^\circ})$, $(\hat{S}_{A^\prime,90^\circ}, \hat{S}_{B^\prime,90^\circ})$ and $(\hat{S}_{A^\prime,45^\circ}, \hat{S}_{B^\prime,45^\circ})$.\newline
The photo-current of the Stokes measurement is down-mixed with an electric local oscillator provided by a function generator. It is the same electric sinusoidal signal with a frequency of 18.2\,MHz as the voltage applied to the EOM. We measure the displacement of the quantum state defined at the sideband frequency of 18.2\,MHz.

The amplified, down-mixed signal is low pass filtered with 2.5\,MHz, sampled by an analog-to-digital converter with 10\,Msamples/s and digitally averaged with 10 samples. Differently displaced modes are obtained by choosing different voltage amplitudes. A Gaussian mixed state is generated by combining this data appropriately on the computer, which is possible because of the ergodicity of the problem. The demodulation for the entanglement recovery is performed computationally as well.
\subsection{Imperfections}

\subsubsection{Experimental Errors}
There are two main sources of experimental errors. First, there is a statistical error in the measurement of the Stokes observables performed for the determination of the covariance matrices and Duan's separability criterion. Second, the shot noise calibration is of limited accuracy. To analyze these errors, the results of the measurements performed on pure coherent states are compared to the theoretical expectations. The information about their discrepancy is used to estimate the error in the further determined values, for example the quantum discord, via a Monte-Carlo simulation. All errors given explicitly in the main text or shown as error bars in the plots, reflect both the error in the calibration as well as the above mentioned statistical error. The error for the covariance matrix given in Eq.\,3 of the main text was estimated to be
\begin{align}
\left(
\begin{array}{cccc}
0.05 & 0.02 & 0.03 & 0.01\\
0.02 & 0.17 & 0.01 & 0.15\\
0.03 & 0.01 & 0.04 & 0.02 \\
0.01 & 0.15 & 0.02 & 0.16
\end{array}
\right).
\end{align}

We must assume that an additional systematic error is present due to further imperfections in the measurement system, as well as due to the modulation performed by the EOM and drifts in the setup over the long measurement times. As a result, the elements of the covariance matrix deviate from what we would expect theoretically. This also applies to the eigenvalues of the covariance matrices. Both for the states originating from initially coherent modes and from modulated squeezed state, we found the lowest eigenvalues less than 1 which would indicate weak global squeezing in the two-mode state after the BS (see Ref. [31] in the main text). But the state prepared by displacing coherent states is  separable by construction. Moreover, also the squeezing in the mixed state prepared from squeezed modes is reliably destroyed due to the amount of imprinted modulation. Thus, the eigenvalue lower than 1 is clearly an artifact of a systematic error present in the setup. This is a common problem of the experimental reconstruction of covariance matrices~\cite{sq_explanation_sup}.


\subsubsection{Common Mode Rejection}
Imperfect common mode
rejection (CMR) in our homodyne detectors is modeled, similar as in~\cite{andersen_sup}, as an addition of a
CM $\gamma_{A'B'}^{CMR}$ to the CM $\gamma_{A'B'}$ of the measured state,
\begin{equation}
\gamma_{A'B'}=\gamma_{A'B'}^{0}+\gamma^{\rm CMR}_{A'B'}
\end{equation}
where $\gamma_{A'B'}^{0}$ is the theoretical model taking into account all other previously mentioned losses and imperfections i.e., BS ratio and attenuation. The CMR addition $\gamma^{\rm CMR}_{A'B'}$ is of the form
\begin{equation}
\gamma_{A'B'}^{\rm CMR}=
\begin{pmatrix}
a&0&0&0\\
0&a&0&0\\
0&0&\tau^2 a& 0\\
0&0 &0& \tau^2 a
\end{pmatrix}, \quad \tau^2+\rho^2=1,
\end{equation}
with $\tau$ the transmittivity of a variable beamsplitter depicted as $\hat{\Lambda}$ in Fig.~1 and $a$ is the additional variance caused by the imperfect common mode rejection and was used as free parameter in the fit. The initial variance introduced by CMR on modes $A$ and $B$ corresponds to $a=3.9\times10^{-3}$ and $a=0.047$, for coherent and squeezed input states respectively. Note that before any local loss is introduced there is an equal level of imperfect CMR in both modes. With increasing dissipation in mode $B$ the noise due to CMR will decrease linearly.

\subsection{Entanglement recovery and Duan's Separability Criterion}

When using the randomly displaced squeezed states as an input in Fig.~1, the initial
nonclassicality is not irreversibly lost.  For example, entanglement which would emerge
after the BS if no displacement is performed, can still be
recovered. If we have access to the displacement $\bar{x}$ encoded on mode $\tilde{E}$,
entanglement between modes $A$ and $B$ can be recovered by directly
performing the reverse displacement on mode $B$ to cancel the modulation.
Initially we have a pure squeezed state $A$ with quadratures $\hat{x}_A=e^{-r}\hat{x}_A^{(0)},\;\hat{p}_A=e^{r}\hat{p}_A^{(0)},$ and mode $B$ in a vacuum state with quadratures $\hat{x}_B=\hat{x}_B^{(0)},\;\hat{p}_B=\hat{p}_B^{(0)}$, $r$ being the squeezing parameter. Random Gaussian displacements  $\bar{x}$ are applied to the $x$ quadrature of mode $A$ such that:
\begin{equation}
\hat{x}_A\rightarrow \hat{x}_A+\bar{x}
\end{equation}
After undergoing a beamsplitter transformation the resulting output quadratures are
\begin{equation}
\begin{split}
\hat{x}'_A=T\hat{x}_A+R\hat{x}_B+T\bar{x},\quad \hat{p}'_A=T\hat{p}_A+R\hat{p}_B,\\
\hat{x}'_B=R\hat{x}_A-T\hat{x}_B+R\bar{x},\quad \hat{p}'_B=R\hat{p}_A-T\hat{p}_B.\\
\end{split}
\end{equation}

The demodulation required for entanglement recovery can be found using the product inseparability criterion \cite{pcrit1_sup, pcrit2_sup}:
\begin{equation}
\langle(g\hat{x}'_A+\hat{x}'_B)^2\rangle \langle(g\hat{p}'_A-\hat{p}'_B)^2\rangle<\frac{1}{4}(g^2+1)^2.
\label{duan}
\end{equation}
The operators on the left-hand side read as
\begin{equation}
\begin{split}
g\hat{x}'_A+\hat{x}'_B&=\hat{x}_A(gT+R)+\hat{x}_B(gR-T)+\bar{x}(gT+R),\\
g\hat{p}'_A-\hat{p}'_B&=\hat{p}_A(gT-R)+\hat{p}_B(gR+T).
\label{parts}
\end{split}
\end{equation}
Hence the general demodulation to be applied to mode $B$ is of the form
\begin{equation} \label{ideal}
\hat{x}''_B\rightarrow\hat{x}'_B-(gT+R)\bar{x},
\end{equation}
which gives,
\begin{equation}
g\hat{x}'_A+\hat{x}'_B=\hat{x}_A(gT+R)+\hat{x}_B(gR-T).
\end{equation}
Rearranging Eq.(\ref{duan}) and using Eq.(\ref{parts}) we get:
\begin{widetext}
\begin{equation}
\frac{[e^{2r}(gT-R)^2+(gR+T)^2]
[e^{-2r}(gT+R)^2+(gR-T)^2]}{(g^2+1)^2}<1.
\label{general}
\end{equation}
\end{widetext}
In the ideal case of a 50:50 beamsplitter $T=R=\frac{1}{\sqrt{2}}$ the left-hand side of the inequality (\ref{general}) is minimised if the gain $g=1$. Therefore
\begin{equation}
e^{-2r}<1,\quad\forall\;r>0,
\end{equation}
and thus entanglement is recovered for any $r>0$.

In this ideal case the demodulation (\ref{ideal}) to be applied to mode $B$ is given by
\begin{equation}
\hat{x}''_B\rightarrow
=\hat{x}'_B-\sqrt{2}\bar{x}.
\end{equation}
Hence the prefactor in the ideal case is $-\sqrt{2}$.

\end{appendix}

\begin{thebibliography}{99}
\bibitem{global_squeezing}
Despite the absence of local squeezing, the CM seems to exhibit a weak global
squeezing. The squeezing occurs in the diagonal direction with respect to
$x$ and $p$ quadratures and thus cannot originate from the input squeezing. We measure the CM of a state, which was created by mixing with vacuum state and therefore lies on the boundary of
the set of squeezed states.
Hence the effect of global squeezing can be attributed to the systematic error caused by
drifts during the long measurement times of the CM and limited balancing between the Stokes measurement setups (see also~\cite{sq_explanation}). These imperfections manifest themselves by the non-zero CM matrix elements refering to $x-p$-correlations, which in turn lead to the artefact of global squeezing.
\end{thebibliography}
\end{document}